\documentclass{elsart}
\usepackage{graphicx}
\usepackage{amssymb}
\begin{document}
\begin{frontmatter}
 \title{Modelling of self-driven particles:\\ foraging ants and pedestrians}
 \author{Katsuhiro Nishinari}
 \address{Department of Aeronautics and Astronautics, University of
 Tokyo, Bunkyo-ku, Tokyo 113-8656, Japan}
 \author{Ken Sugawara}
 \address{Department of Information Science,
Tohoku Gakuin University, Sendai, 981-3193, Japan}
\author{Toshiya Kazama}
\address{Department of Computational Intelligence and Systems Science, 
Tokyo Institute of Technology, meguro-ku, Tokyo 152-8552, Japan}
 \author{Andreas Schadschneider}
 \address{Institut  f\"ur Theoretische  Physik, Universit\"at 
zu K\"oln D-50937 K\"oln, Germany}
 \author{Debashish Chowdhury}
 \address{Department of Physics, Indian Institute of Technology,
Kanpur 208016, India}
\begin{abstract}
  Models for the behavior of ants and pedestrians are studied in an
  unified way in this paper.  Each ant follows pheromone put by
  preceding ants, hence creating a trail on the ground, while
  pedestrians also try to follow others in a crowd for efficient and
  safe walking.  These following behaviors are incorporated in our
  stochastic models by using only local update rules for computational
  efficiency.  It is demonstrated that the ant trail model shows an unusual
  non-monotonic dependence of the average speed of the ants on their
  density, which can be well analyzed by the zero-range process.  We
  also show that this anomalous behavior is clearly observed in an
  experiment of multiple robots.  Next, the relation between the ant
  trail model and the floor field model for studying evacuation
  dynamics of pedestrians is discussed.  The latter is
  regarded as a two-dimensional generalization of the ant trail model,
  where the pheromone is replaced by footprints.  It is shown from
  simulations that small perturbations to pedestrians will sometimes
  avoid congestion and hence allow safe evacuation.
\end{abstract}
 \end{frontmatter}
\section{Introduction}
Modelling of self-driven particles is a recently progressing research
area in physics, and opens a new paradigm in statistical physics
\cite{helbing,css}.  Self-driven particles, such as vehicles,
pedestrian and ants, do not necessarily satisfy Newton's third law,
i.e., the law of action and reaction.  This is mainly because the
force acting among such particles has a psychological or social
origin, and hence the reaction differs from one individual to another.
Therefore the behavior of these particles is not well described by the
usual framework of mechanics or (equilibrium) statistical physics.
Nevertheless in the case of collective motion, some aspects of the
behavior have similarities with those seen in Newtonian particles, such
as phase transitions, cluster formation and occurrence of domain walls
\cite{helbing,css}.

There are several attempts to describe such self-driven particles for
the last several decades, ranging from differential equations with
some social forces \cite{HFV,DFHD}, to cellular automaton (CA) models
with various update rules \cite{NS,css}.  In this paper, we show
modellings of ants and pedestrians by a cellular automaton approach in
an unified way.  We demonstrate similarities between ants on a trail
and pedestrians in evacuations, since both try to follow others through
sniffing pheromone on a trail, and information from one's eyes and
ears, respectively.  The behavior of pedestrians is usually not simple
in normal situations, but in emergency cases like evacuation from a
building on fire, people rush to exits and only simple motions are
observed.  In this paper we restrict ourselves to the case of
evacuation in the modelling of pedestrians. For this case, we show
that human behavior can be modeled by a naturally extended model of
ants.

The paper is organized as follows. In Sec.~2, we introduce our model
of traffic of ants on a trail, and discuss its dynamical properties
such as loose cluster formation and anomalous flow-density relation by
using a solvable stochastic model.  We have also performed an
experiment on our ant trail model by developing a multiple robots
system.  In Sec.~3, a stochastic model of pedestrian dynamics is
presented, which is called the {\it floor field} model.  We discuss
the relation of the floor field model and the ant trail model, and
show simulated results of the model by taking into account the inertia
of pedestrians.  Concluding remarks and future problems are given in
Sec.~4.

\section{Traffic of ants}
\subsection{Ant trail model}\label{sec2}

Let us first define the ant trail model (ATM) which is a simple model for
an unidirectional motion of ants on a trail.  The ants communicate
with each other by dropping a chemical called {\it pheromone} on the
substrate as they move forward \cite{mc}.  The pheromone sticks to the
substrate long enough for the other following ants to pick up its
smell and follow the trail.

First we divide the one-dimensional trail into $L$ cells, each of
which can accommodate at most one ant at a time.  We will use
periodic boundary conditions, that is, we consider an ant
trail of a circuit type.  The lattice cells are labeled by the index
$i$ ($i = 1,2,...,L$), and we associate two binary variables $S_i$ and
$\sigma_i$ with each site $i$.  $S_i$ takes the value $0$ or $1$
depending on whether the cell is empty or occupied by an ant.
Similarly, $\sigma_i = 1$ if the cell $i$ contains pheromone;
otherwise, $\sigma_i = 0$.  Since a unidirectional motion is assumed,
ants do not move backward in this model.  Their hopping probability
becomes higher if it smells pheromone ahead of it.  The state of
the system is updated at each time step in two stages.  In stage I,
ants are allowed to move, and stage II corresponds to the evaporation
of pheromone.  In each stage the dynamical rules are applied in
parallel to all ants and pheromones, respectively.

\noindent {\it Stage I: Motion of ants}\\[0.2cm]
\noindent An ant at cell $i$ that has an empty cell in front of it, i.e., 
$S_i(t)=1$ and $S_{i+1}(t)=0$, hops forward with probability $Q$
if $\sigma_{i+1}(t) = 1$ and $q$ if $\sigma_{i+1}(t) = 0$.
Here we assume $ q < Q$ for consistency with real ant-trails where
pheromone works as an attractive chemical for ants on a trail.

\noindent {\it Stage II: Evaporation of pheromones}\\[0.2cm]
\noindent At each cell $i$ occupied by an ant after stage I
a pheromone will be created, i.e., 
$\sigma_i(t+1) = 1$ if $S_i(t+1) = 1$.
On the other hand, any free pheromone at a site $i$ not occupied
by an ant will evaporate with the probability $f$ 
per unit time, i.e., if $S_i(t+1) = 0$, $\sigma_i(t) = 1$, then
$\sigma_i(t+1) = 0$ with probability $f$ and $\sigma_i(t+1) = 1$
with probability $1-f$.

The rules can be written in a compact form as 
the following coupled equations:
\begin{eqnarray}
 S_j(t+1)&=&S_j(t)+\min(\eta_{j-1}(t),S_{j-1}(t),1-S_j(t))\nonumber\\ 
&&\hspace{0.5cm}-\min(\eta_{j}(t),S_{j}(t),1-S_{j+1}(t)),\label{eqa}\\
\sigma_j(t+1)&=&\max(S_j(t+1),\min(\sigma_j(t),\xi_j(t))),\label{eqf}
\end{eqnarray}
where $\xi$ and $\eta$ are stochastic variables defined by
$\xi_j(t)=0$ with probability $f$ and $\xi_j(t)=1$ with probability
$1-f$, and $\eta_j(t)=1$ with probability $p=q+(Q-q)\sigma_{j+1}(t)$ and
$\eta_j(t)=0$ with probability $1-p$.  We point out that eqs.~(\ref{eqa}) and
(\ref{eqf}) reduce to the asymmetric simple exclusion process (ASEP)
\cite{ASEP}, which is one of the exactly solvable models, if 
$p$ is a constant, i.e., $p$ does not depend on $\sigma$. If we
further consider the deterministic limit $p=1$, then this model
reduces to the Burgers CA \cite{NT}, which is also exactly solvable.

\subsection{Numerical results}
The ASEP with parallel updating has been used often as a simple model
of vehicular traffic on single-lane highways \cite{NS}.  The most
important quantity of interest in the context of flow properties of
the traffic models is the {\it fundamental diagram}, i.e., the
flow-versus-density relation, where flow is the product of the density
$\rho$ and the average speed $V$.  Thus it is interesting to compare the
fundamental diagram of the ATM with that for vehicular traffic.
\begin{figure}[hb]
\begin{center}
\includegraphics[width=0.44\textwidth]{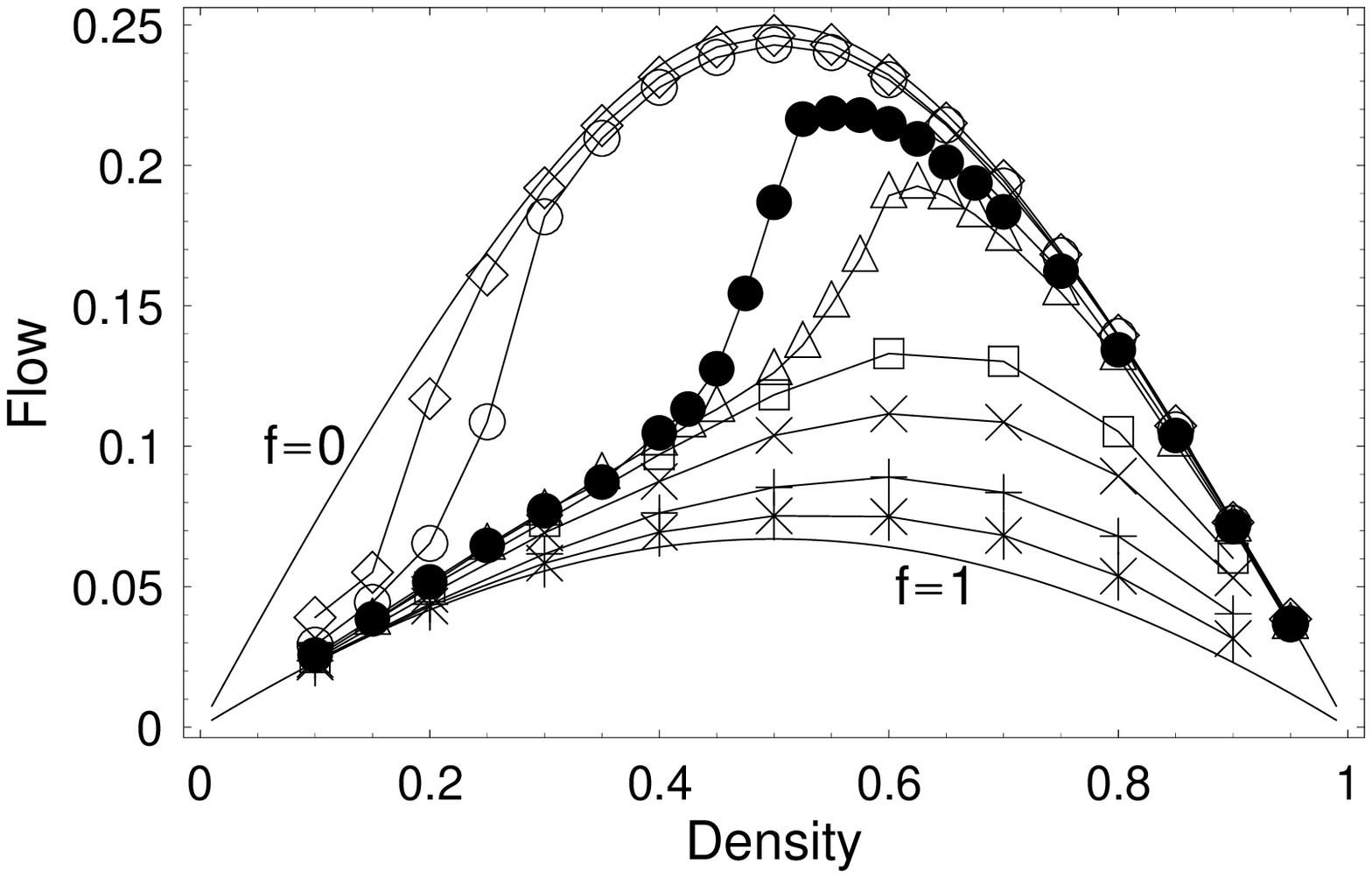}
\includegraphics[width=0.44\textwidth]{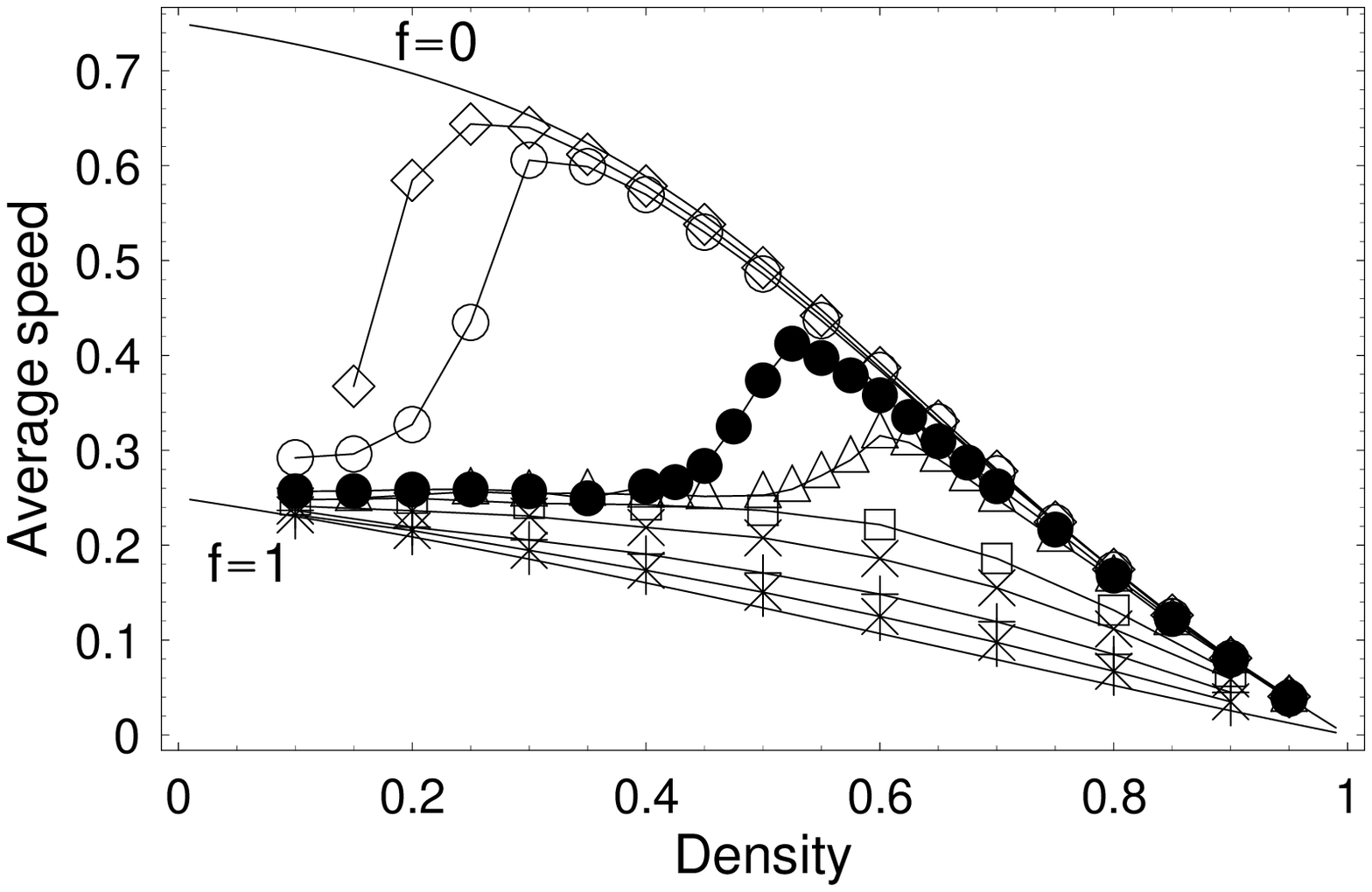}\\
(a)\hspace{5cm}(b)
\end{center}
\caption{The average flow (a), speed (b) of the ants are plotted against
  their densities for the parameters $Q = 0.75, q = 0.25$ and $L=500$.
  The discrete data points correspond to $f=0.0005 ({\Diamond})$,
  $0.001 (\circ$), $0.005 (\bullet)$, $0.01 ({\bigtriangleup})$, $0.05
  ({\Box})$, $0.10 (\times)$, $0.25 (+)$, $0.50 (\ast)$.  The cases
  $f=0$ and $f=1$ are also displayed, which are identical to the ASEP
  corresponding to the effective hopping probabilities $Q$ and $q$,
  respectively.}
\label{fig-effh}
\end{figure}

Numerical results of the fundamental diagrams of the ATM is given in
Fig.~\ref{fig-effh}(a).  The density-dependence of the average speed
of the ATM is also shown in Fig.~\ref{fig-effh}(b).
We first observe that the diagram does not possess particle-hole
symmetry.  In the ASEP the flow remains invariant under the
interchange of $\rho$ and $1-\rho$; this particle-hole symmetry leads
to a fundamental diagram that is symmetrical about $\rho =
\frac{1}{2}$.  In the ATM, particle-hole symmetry is seen in the
special cases of $f = 0$ and $f = 1$ from Fig.~\ref{fig-effh}(a).
In these two special cases the ant-trail model becomes identical to the 
ASEP with parallel updating corresponding to the effective hopping 
probabilities $Q$ (for $f = 0$) and $q$ (for $f = 1$),
respectively.

Next, over a range of small values of $f$, it exhibits an anomalous
behavior in the sense that, unlike common vehicular traffic, $V$ is
not a monotonically decreasing function of the density $\rho$
(Fig.~\ref{fig-effh}(b)).  Instead a relatively sharp crossover can be
observed where the speed {\em increases} with the density. In the
usual form of the fundamental diagram (flow vs.\ density) this
transition leads to the existence of an inflection point
(Fig.~\ref{fig-effh}(a)).

\begin{figure}[htb]
\begin{center}
\includegraphics[width=0.45\textwidth]{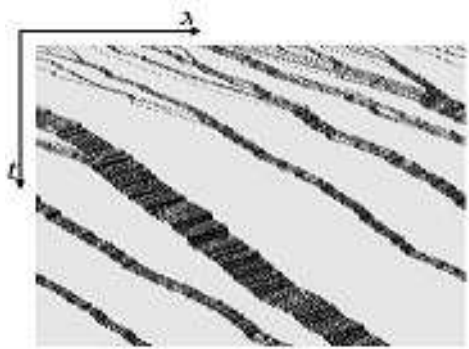}
\end{center}
\caption{Spatial-temporal behaviors of loose clusters 
in the low density case ($\rho=0.16$). Parameters
are $Q = 0.75, q = 0.25, f=0.005$. We see the loose clusters emerge
 from the random initial configuration, which will eventually merge
after sufficiently long time into one big loose cluster.}
\label{loose}
\end{figure}

By a detailed study of the spatio-temporal behavior in the
steady-state, we were able to distinguish three different density
regimes \cite{nds}.  At low densities a loosely assembled cluster is
formed and propagates with the probability $q$ (Fig.~\ref{loose}).
The leading ant in the cluster which hops with probability $q$ will
determine the velocity of the cluster.  Homogeneous mean field
theories fail in this region since these theories always assume that
ants are uniformly distributed, which is not true in the case of the
ATM \cite{nds}.

In the intermediate density regime, the leading ant occasionally hops
with probability $Q$ instead of $q$, because it becomes possible to
feel the pheromone which is dropped by the last ant in the preceding
cluster.  This arises from the fact that, due to the periodic boundary
conditions, the gap size between the last and leading ant becomes
shorter as the cluster becomes larger, so that the leading ant is
likely to find the pheromone in front of it. This increase of the
average speed in this intermediate-density region
(Fig.~\ref{fig-effh}(b)), leads to the anomalous fundamental diagram.

Finally, at high densities, the mutual hindrance of the ants dominates
the flow behavior leading to a homogeneous state similar to that of
the ASEP. In this regime the loose cluster does no longer exist and
ants are uniformly distributed after a long time.  Thus homogeneous
mean field theories give a good result in accounting the fundamental
diagram only in this regime \cite{nds}.

\subsection{Analytical results}
\label{subanalytic}
\begin{figure}[htb]
\begin{center}
\includegraphics[width=0.55\textwidth]{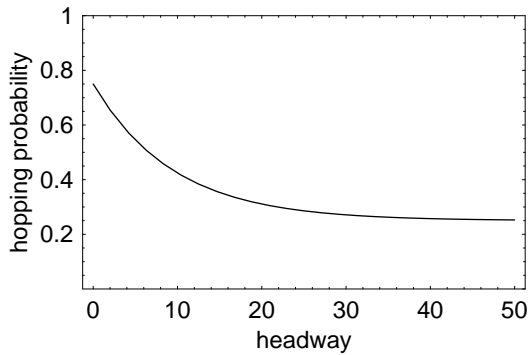}
\end{center}
\caption{Hopping probability versus the number of cells between successive
 ants. Parameters are $Q=0.75,q=0.25,f=0.1$ and $V=1$ in (\ref{hop}).}
\label{hopant}
\end{figure}

The ATM is closely related to the zero-range process (ZRP), which is
known as one of the exactly solvable models of interacting
Markov processes \cite{spi,eh-review}.
The ZRP consists of the moving particles of the exclusion process.
The hopping probability of a particle depends on 
the number of cells to the particle in front. 
Hence the ZRP is regarded as a generalization of ASEP.
In the ATM, the hopping probability $u$ can be expressed as
\begin{equation}
 u=q(1-g) + Q g,
\label{hop}
\end{equation}
where $g(x)=(1-f)^{x/V}$ (Fig.~\ref{hopant}).
Assuming that a gap size of successive ants is $x$,
$g$ represents the probability
that there is a surviving pheromone 
on the first site of a gap \cite{cgns}.
Thus, since in the ATM the hopping probability $u$ is related to $x$,
we may utilize the exact results for the ZRP \cite{eh-review}. The
average velocity $V$ of ants is then calculated by
\begin{equation}
 V=\sum_{x=1}^{L-M}u(x)p(x)
\label{veldef}
\end{equation}
where $L$ and $M$ are the system size and the number of ants 
respectively (hence $\rho=M/L$ is the density), and $p(x)$
is the probability of finding a gap of size $x$, which is given by 
\begin{eqnarray}
p(x) = h(x) \frac{Z(L-x-1,M-1)}{Z(L,M)}.
\label{ph}
\end{eqnarray}
Since the ATM is formulated with parallel update, the form of $h(x)$, 
as calculated in (\ref{ph}), is given by \cite{Evans97}
\begin{equation}
 h(x)=\left\{\begin{array}{cc}
1-u(1) & \qquad{\rm for} \,\,\,\,\,\,\, x=0\phantom{\ .}\\
\displaystyle{\frac{1-u(1)}{1-u(x)}\prod_{y=1}^x \frac{1-u(y)}{u(y)}}
& \qquad{\rm for} \,\,\,\,\,\,\, x>0\ .
\end{array}
\right.
\end{equation}
The partition function $Z$ is obtained by the recurrence relation
\begin{equation}
 Z(L,M)=\sum_{x=0}^{L-M}Z(L-x-1, M-1)h(x),
\end{equation}
with $Z(x,1)=h(x-1)$ and $Z(x,x)=\{h(0)\}^x$, which is easily obtained
by (\ref{ph}) with the normalization $\sum p(x)=1$.

\begin{figure}[htb]
\begin{center}
\includegraphics[width=0.55\textwidth]{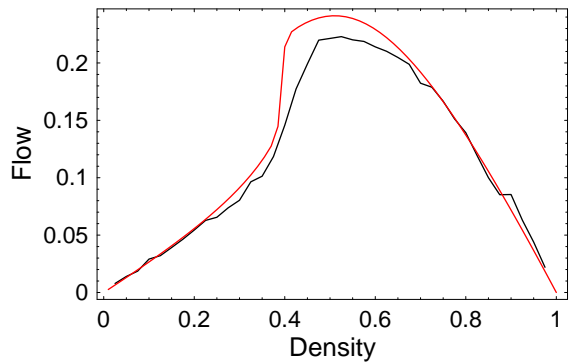}
\caption{The fundamental diagram of the ATM with the parameter $L=200$. 
Parameters are $Q=0.75, q=0.25,f=0.005$.
The smooth thin curve is the theoretical one, while
the zigzaged thick one is the numerical data.}
\end{center}
\label{zrpfund}
\end{figure}
Next we draw the fundamental diagram of the ATM by using
(\ref{veldef}) and changing $\rho$ from 0 to 1.  It is given in
Fig.~4 with $L=200$.  The thick black curve is the
numerical data and the smooth thin black one is the theoretical curve
in each figure with the specified value of $L$.  We see that the
theoretical curves are almost identical to the numerical ones, thus
confirming that the steady state of the ATM is described by the ZRP.

\subsection{Multiple robots experiment}
Next we set up an experimental system of our ant model by using
multiple robots that can communicate with each other through
pheromone-like interaction.  In this experiment, a virtual pheromone
system (V-DEAR) \cite{suga} was introduced, in which chemical signals
are mimicked by the graphics projected on the floor, and the robots
decide their action depending on the color information of the
graphics.

\begin{figure}[htb]
\begin{center}
\includegraphics[width=0.9\textwidth]{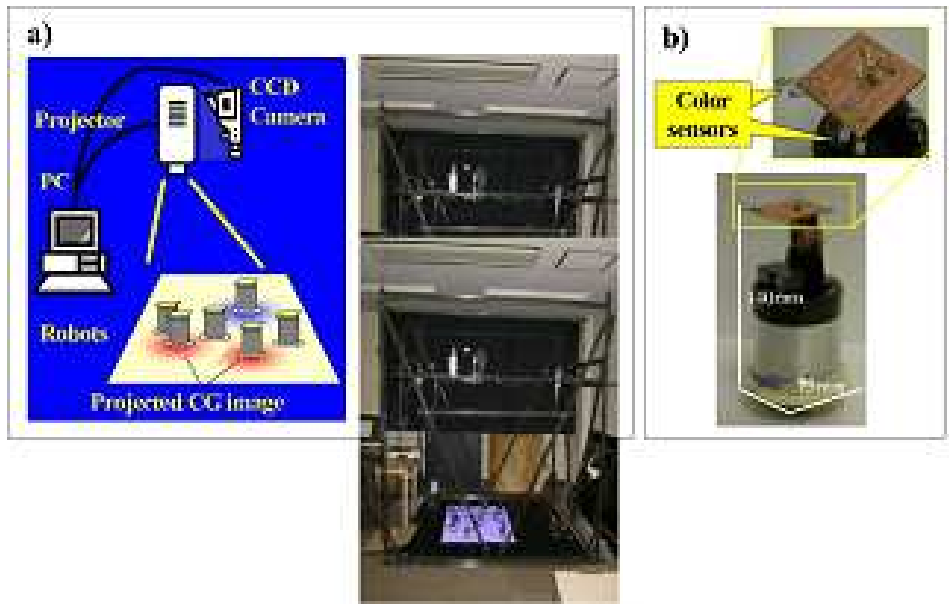}
\caption{A schematic view and photo of V-DEAR. 
(a) The system is composed of Liquid Crystal Projector to 
project the Computer Graphics 
and the CCD camera to trace the position of the robots in the field. 
(b) Each robot has sensors on the 
top to detect the color and brightness of the field.}
\end{center}
\label{robotsystem}
\end{figure}

Fig.~5 shows the schematic and photo of V-DEAR.  Each
robot moving on the field detects the color and brightness of the
field from sensors on the top, and determines its actions autonomously
based on the conditions of the Computer Graphics (CG) on the floor.
Combining the position information of the robots acquired from the CCD
camera and the projected CG by the projector, we can achieve the
dynamic interaction between the environment and robots.

Each robot leaving a pheromone moves in the same direction 
on the circumference of a circle with diameter 
520~mm (Fig.~6, left). 
The robot immediately behind it traces the pheromone, hence it
follows the leader. 
The speed of each robot $v(x)$ is given by 
\begin{equation}
v(x) = \left\{\begin{array}{cc}
 V_Q  &   \hspace{4mm} {\rm for}\hspace{2mm} p(x, t) \ge p_{\rm th} \\
 V_q  &  \hspace{4mm} {\rm for} \hspace{2mm} p(x, t) <  p_{\rm th} 
\end{array}
\right.
\label{robov}
 \end{equation}
where $x$ is the position on the circumference, 
$p(x,t)$ is the concentration of pheromone at the position 
and $p_{\rm th}$ is a constant (Fig.~6, middle). 
If a robot catches up with the robot in front, then 
the robot in the rear stops moving for 1~sec (Fig.~6, right). 

\begin{figure}[htb]
\begin{center}
\includegraphics[width=0.9\textwidth]{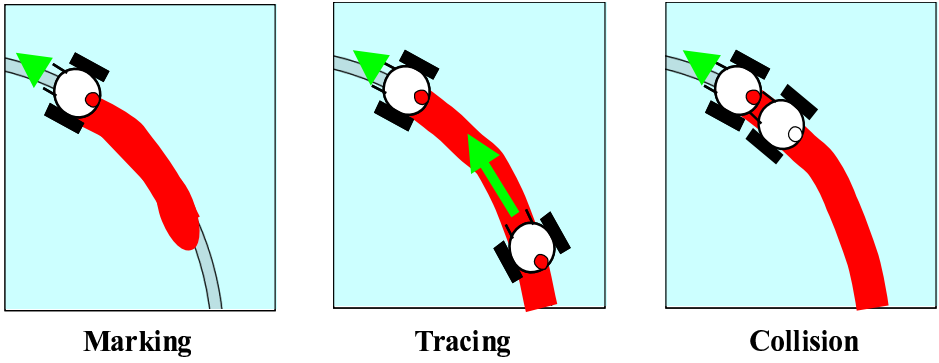}
\caption{Interactions between robots through the color CG projected on
 the floor. Marking, tracing and collision
are schematically depicted. 
}
\end{center}
\label{robotint}
\end{figure}

The dynamics of the virtual pheromone is described as
\begin{equation}
 \frac{dp(x, t)}{dt} = P - f p(x, t), 
\end{equation}
where $f$ is the rate of evaporation and 
$P$ the injection concentration given by
$P=P_0$ if a robot exists at $x$, and $P=0$ if not.
$p(x, t)$ is expressed by the 
gradation of the white color CG which ranges from 0 to 255. 
In this experiment, we set
$V_Q =7$~cm/s, $V_q=1.4$~cm/s, $p_{\rm th}=127$, and $P_0=250$.

The experiment was carried out with groups of 4 to 17 robot 
teams, with $f=0$, 0.03, 0.05 and 1.00. 
All experiments were performed one time and had a duration of 3 minutes. 
As initial condition, robots were distributed evenly on the circle. 
\begin{figure}[htb]
\begin{center}
\includegraphics[width=1\textwidth]{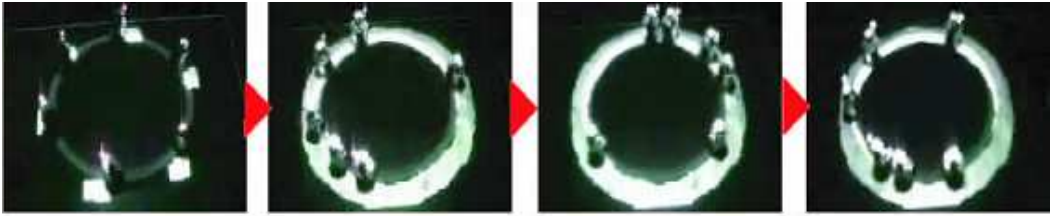}
\caption{Loose cluster formation by six robots on a circle.}
\end{center}
\label{robotmarch}
\end{figure}

Fig.~7 shows an example of loose cluster formation
of robots in this experiment.  The maximal group size of robots on the
circumference is 22, hence Fig.~7 corresponds to a
density of 6/22.  Flow is defined as
the number of robots crossing a fixed point on the circumference
during the experimental duration.  The fundamental diagram of our
experiment is given in Fig.~8.
\begin{figure}[htb]
\begin{center}
\includegraphics[width=0.48\textwidth]{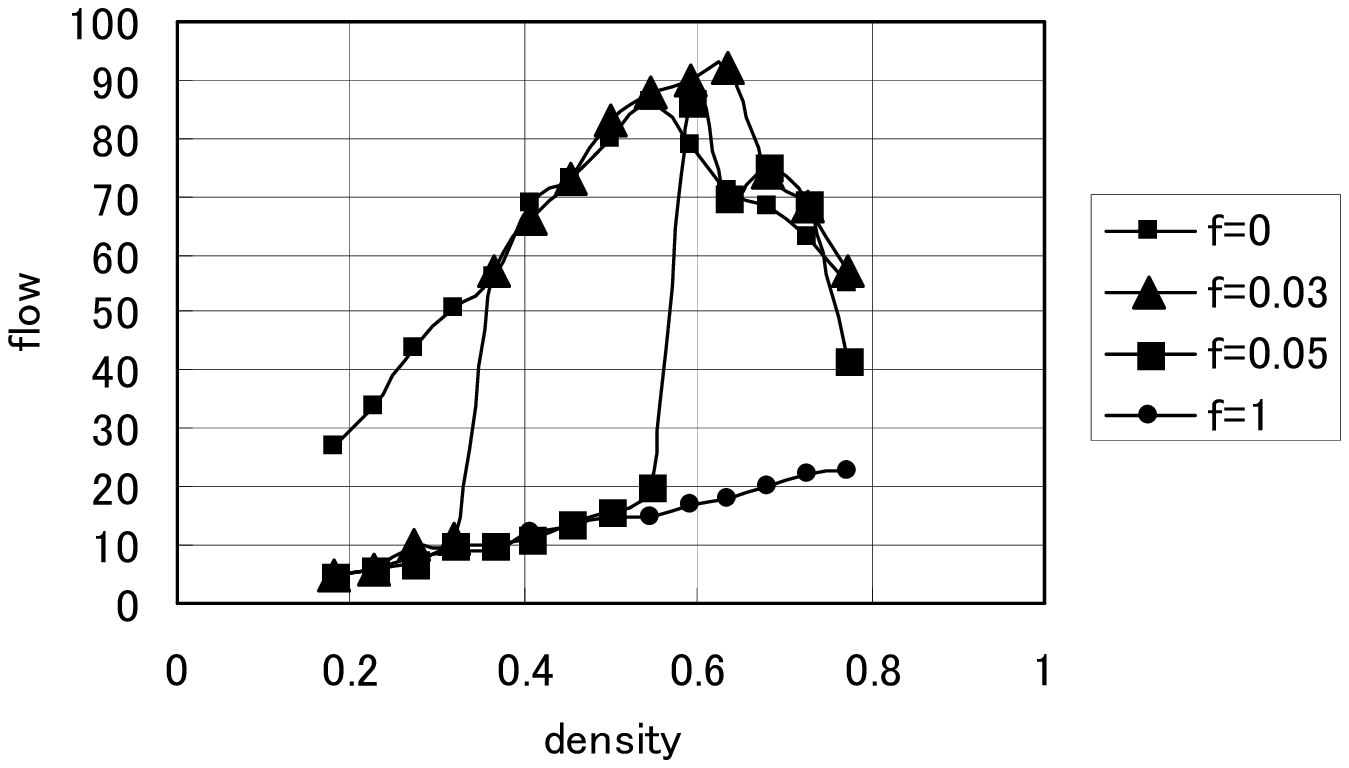}
\includegraphics[width=0.43\textwidth]{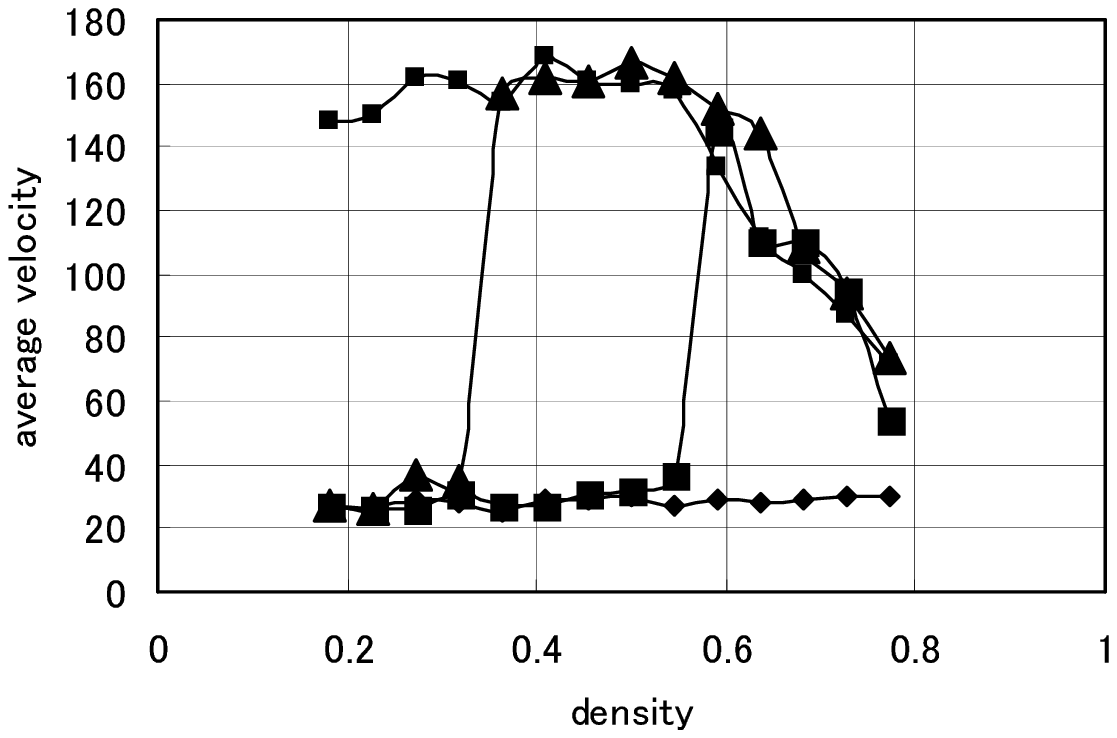}
\caption{Fundamental diagram of the robots experiments for
$f=0, 0.03, 0.05$ and 1. We see that the velocity increases sharply
for the small but finite values of $f$, which is similar to fig.1.}
\end{center}
\label{robotfund}
\end{figure}
It shows anomalous features  very similar to
Fig.~\ref{fig-effh}, although the robot experiment is subject to
various perturbations.  From Fig.~8 the average velocity surely
increases only in the small $f$ cases.  Thus we confirm in this
experiment that our finding on the anomalous behavior in the ATM is
robust and may happen in nature if the evaporation rate of pheromone
is sufficiently low.

\section{Modelling of evacuations}
\subsection{Floor field model}
We consider a model for pedestrians in evacuation in this section by
using a cellular automaton approach.  The model is called the floor
field model \cite{BKAJ} (FFM), which is successfully used up to now
for modelling panic behavior of people evacuating from a room.  The
two-dimensional space is discretized into cells of size $40~{\rm cm}
\times 40~{\rm cm}$ which can either be empty or occupied by one
pedestrian.  Each pedestrian can move to one of the unoccupied
Neumann-neighbor cells $(i,j)$ or stay at the present cell at each
time step $t\to t+1$ according to certain transition probabilities
$p_{ij}$ as explained below.  The important step for this model is to
introduce two kinds of floor fields: The {\em static floor field} $S$
describes the shortest distance to an exit door. The field strength
$S_{ij}$ is set inversely proportional to the distance from the door.
The other one is the {\em dynamic floor field} $D$, which is a
virtual trace left by the pedestrians similar to the pheromone in
chemotaxis.  It has its own dynamics, namely diffusion and decay,
which leads to broadening, dilution and finally vanishing of the
trace.

The update rules of the FFM have the following structure:
\begin{enumerate}
\item The dynamic floor field $D$ is modified according to its
diffusion and decay rules, controlled by the parameters $\alpha$
and $\delta$. 

\item For each pedestrian, the transition probabilities $p_{ij}$
for a move to an unoccupied neighbor cell $(i,j)$ are determined
by the two floor fields.
The values of the fields $D$ (dynamic) and $S$ (static) are
weighted with two sensitivity parameters $k_D$ and $k_S$:
\begin{equation}
  p_{ij} \sim \exp{\left(k_D D_{ij}\right)}
  \exp{\left(k_S S_{ij}\right)}P_I(i,j)\xi_{ij}.
\label{trap}
\end{equation}
Here $P_I$ represents the inertia
effect \cite{ICnishi} given by $P_I(i,j)=\exp{\left(k_I \right)}$ for
the direction of one's motion in the previous time step, and
$P_I(i,j)=1$ for other cells, where $k_I$ is the sensitivity
parameter. 
$\xi_{ij}$ is responsible for the exclusion in a cell; if the target
cell is occupied by another pedestrian or an obstacle, then we set
$\xi_{ij}=0$, while $\xi_{ij}=1$ if it is free.

\item Each pedestrian chooses randomly a target 
cell based on the transition
      probabilities $p_{ij}$ determined by (\ref{trap}).

\item Whenever two or more pedestrians attempt to move to the
      same target cell, the movement of {\em all}
involved particles is denied with probability $\mu \in [0,1]$,
i.e.\ all pedestrians remain at their site. 
Which one is allowed to move is decided using a
probabilistic method \cite{AKA}.

\item The pedestrians who are allowed to move perform their motion
to the target cell chosen in step 3. $D$ at the origin cell
$(i,j)$ of each {\em moving} particle is increased by one:
$D_{ij}\to D_{ij}+1$.

\end{enumerate}
The above rules are applied to all pedestrians at the same time
(parallel update). 

Let us now briefly explain the similarities between the FFM and the ATM.
Considering a one-dimensional variant of the FFM,
the hopping probability (\ref{trap}) of moving forward
for a pedestrian at cell $i$ is simply written as
$p_i \sim \exp(k_D D_{i+1})$,
where $D_{i+1}$ is the number of footprints at the next cell $i+1$.
Here assuming a unidirectional motion of pedestrians, 
the static floor field and inertia can be omitted in (\ref{trap}).
Thus the FFM takes into account the
concentration of multiple pheromones, i.e., footprints at a cell.
In the ATM, $p_i$ takes only two values, $Q$ and $q$, depending on 
$D_{i+1}=1$ or 0 respectively. Thus if we restrict the binary values for
$D$ as $D\to\min(1,D)$, then the one-dimensional FFM
directly corresponds to the ATM 
if we choose the hopping probability as
\begin{equation}
p_i=q\exp(k_D \min(1,D_{i+1}))
\end{equation}
where $k_D=\ln Q/q$. 
\subsection{Model parameters and their physical relevance}
There are several parameters in the FFM which are
listed below with their physical meanings.
\begin{enumerate}
\item $k_S\in [0,\infty ) \cdots $ The coupling to the static field
characterizes the knowledge of the shortest path to the doors, or
the tendency to minimize the costs due to deviation from a planned
route. 
 \item $k_D\in [0,\infty ) \cdots $ The coupling to the dynamic
field characterizes the tendency to follow other people in motion ({\em
herding behavior}). The ratio $k_D/k_S$ may be interpreted as the
degree of panic. It is known that people try to follow others
particularly in panic situations \cite{HFV}. 

 \item $k_I\in [0,\infty ) \cdots $ This parameter determines the
 strength of inertia which suppresses quick changes of the 
direction of motion.
It also reflects the individual's tendency to avoid unnecessary
acceleration of the speed during walking.

 \item $\mu\in[0,1]\cdots $ The friction parameter controls
 the resolution of conflicts in clogging situations. Both
cooperative and competitive behavior at a bottleneck are well
described by adjusting  $\mu$ \cite{AHKAM}.

 \item $\alpha, \delta\in[0,1]\cdots $ These constants control
diffusion and decay of the dynamic floor field. It reflects the
randomness of people's movement and the visible range of a person,
respectively. If the room is full of smoke, then $\delta$ takes
large value due to the reduced visibility. 

\end{enumerate}

\subsection{Simulations}\label{simulation}
\begin{figure}[htb]
\begin{center}
\includegraphics[width=0.55\textwidth]{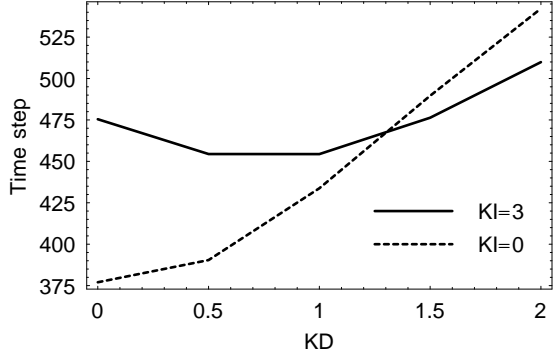}
\end{center}
\caption{
Total evacuation time vs.\ coupling $k_D$ to the dynamic
floor field for different values of $k_I$.
The room is a simple square without obstacles and 50 simulations are
averaged for each data point.
Parameters are $\rho=0.04, k_S=2$, $\alpha=0.2$, $\delta=0.2$ and $\mu=0$.}
\label{figinertia}
\end{figure}

We focus on measuring the total evacuation time by changing the
parameter $k_I$.  The size of the room is set to 100 $\times$ 100
cells.  Pedestrians try to keep their preferred velocity and direction
as long as possible for minimizing their effort.  This is taken into
account by adjusting the parameter $k_I$. In Fig.~\ref{figinertia},
total evacuation times from a room without any obstacles are shown as
function of $k_D$ in the cases $k_I=0$ and $k_I=3$.  We see that it is
monotonously increasing in the case $k_I=0$.  This is because any
perturbation from other people becomes large if $k_D$ increases, which
causes the deviation from the minimum route.  Introduction of inertia
effects, however, changes this property qualitatively, and the {\it
  minimum} time appears around $k_D=1$ in the case $k_I=3$.  This is
well explained by taking into account the physical meanings of $k_I$
and $k_D$.  If $k_I$ becomes larger, people become less flexible and
all of them try to keep their own minimum route to the exit according
to the static floor field regardless of congestion.  For the
(unrealistic) case of very large values of $k_I$ their motion will be
completely determined by inertia and thus they might not be able
to find the exit at all. By increasing
$k_D$, disturbances from other people become relevant through the
dynamic floor field.  This perturbation makes one flexible and hence
contributes to avoid congestion, especially at the exit.  Large $k_D$
again works as strong perturbation as in the case of $k_I=0$, which
diverts people from the shortest route largely.  Note that if we take
sufficiently large $k_I$, then the effect of $k_D$ becomes relatively
weak and smooth evacuation will be prevented due to the strong
clogging at the exit.

\section{Conclusion}\label{secconcl}
In this paper, we have studied simple models for ants and pedestrians
in an unified way.  Ants on a trail follows pheromone dropped by its
preceding ants, and pedestrians also follow others in a crowded or
panic situation mainly for safe and efficient reasons.  We have
proposed stochastic cellular automaton models of both ants on a trail
and pedestrians in evacuation.

The ant trail model shows an anomalous flow-density relation, which
can be analyzed by a ZRP.  We have shown that this behavior is also
seen in an experiment of multiple robots.  It is interesting to see
whether the fundamental diagrams remains anomalous when we introduce
more then two velocities in (\ref{robov}). Since the velocity of ants
in reality allows more gradations, it is important to test how robust
the anomaly predicted by the simple model is.

Some generalizations of the ATM have been recently studied: For open
boundary conditions, we have a similar phase diagram where the
critical point depends on the evaporation probability $f$ \cite{jpsj}.
A simple bidirectional generalization is also proposed in \cite{bi}.
It is interesting to compare the fundamental diagram recently obtained
in an experiment with bidirectional leaf-cutting ants \cite{burd} with
that of the bidirectional model.  The FFM can be regarded as a
two-dimensional generalization of the ATM, where the pheromone is
replaced by footprints.  By using the FFM it is shown that the small
perturbation to pedestrians can reduce the total evacuation time in
some cases.

Finally, it is interesting to study how a trail of ants between their
nest and a food source is created by extending the ATM to two spatial
dimension.  Then we can directly compare the behaviors of ants and
pedestrians by using the two models, and hopefully we will find good
strategies of evacuation from the behaviors of ants that have ``swarm
intelligence'' \cite{swarmint}.


\end{document}